\documentclass[12pt,a4paper]{article}
\usepackage{ifthen} % for conditional statements
\newboolean{pdflatex}
\setboolean{pdflatex}{false} % False for eps figures 

\newboolean{articletitles}
\setboolean{articletitles}{true} % False removes titles in references

\newboolean{uprightparticles}
\setboolean{uprightparticles}{false} %True for upright particle symbols
\usepackage{graphicx}
\usepackage{microtype}
\usepackage{lineno}  % for line numbering during review
\usepackage{xspace} % To avoid problems with missing or double spaces after
\usepackage{graphicx}  % to include figures (can also use other packages)
\usepackage{color}
\usepackage{colortbl}
\graphicspath{{./figs/}} % Make Latex search fig subdir for figures

\usepackage{amsmath} % Adds a large collection of math symbols
\usepackage{amssymb}
\usepackage{amsfonts}
\usepackage{upgreek} % Adds in support for greek letters in roman typeset

\newcommand*\patchAmsMathEnvironmentForLineno[1]{%
\expandafter\let\csname old#1\expandafter\endcsname\csname #1\endcsname
\expandafter\let\csname oldend#1\expandafter\endcsname\csname
end#1\endcsname
 \renewenvironment{#1}%
   {\linenomath\csname old#1\endcsname}%
   {\csname oldend#1\endcsname\endlinenomath}%
}
\newcommand*\patchBothAmsMathEnvironmentsForLineno[1]{%
  \patchAmsMathEnvironmentForLineno{#1}%
  \patchAmsMathEnvironmentForLineno{#1*}%
}
\AtBeginDocument{%
\patchBothAmsMathEnvironmentsForLineno{equation}%
\patchBothAmsMathEnvironmentsForLineno{align}%
\patchBothAmsMathEnvironmentsForLineno{flalign}%
\patchBothAmsMathEnvironmentsForLineno{alignat}%
\patchBothAmsMathEnvironmentsForLineno{gather}%
\patchBothAmsMathEnvironmentsForLineno{multline}%
}

\usepackage{hyperref}    % Hyperlinks in references
\usepackage[all]{hypcap} % Internal hyperlinks to floats.

\def\lhcb {\mbox{LHCb}\xspace}
\def\ux85 {\mbox{UX85}\xspace}

\def\lhc    {\mbox{LHC}\xspace}

\def\cdf    {\mbox{CDF}\xspace}
\def\dzero  {\mbox{D0}\xspace}

\ifthenelse{\boolean{uprightparticles}}%
{

 \def\Pmu         {\ensuremath{\upmu}\xspace}

 \def\Ppi         {\ensuremath{\uppi}\xspace}

 \def\Ppsi        {\ensuremath{\uppsi}\xspace}

 \def\PDelta      {\ensuremath{\Delta}\xspace}                 
 \def\PXi      {\ensuremath{\Xi}\xspace}                 
 \def\PLambda      {\ensuremath{\Lambda}\xspace}                 
 \def\PSigma      {\ensuremath{\Sigma}\xspace}                 
 \def\POmega      {\ensuremath{\Omega}\xspace}                 
 \def\PUpsilon      {\ensuremath{\Upsilon}\xspace}                 
 
 \def\PB      {\ensuremath{\mathrm{B}}\xspace}                 
                  
 \def\PD      {\ensuremath{\mathrm{D}}\xspace}

 \def\PJ      {\ensuremath{\mathrm{J}}\xspace}                 
 \def\PK      {\ensuremath{\mathrm{K}}\xspace}

 \def\Pb      {\ensuremath{\mathrm{b}}\xspace}                 
 \def\Pc      {\ensuremath{\mathrm{c}}\xspace}

 \def\Pi      {\ensuremath{\mathrm{i}}\xspace}

 \def\Ps      {\ensuremath{\mathrm{s}}\xspace}

}
{

 \def\Pmu         {\ensuremath{\mu}\xspace}

 \def\Ppi         {\ensuremath{\pi}\xspace}

 \def\Ppsi        {\ensuremath{\psi}\xspace}                 
                  
 \mathchardef\PDelta="7101
 \mathchardef\PXi="7104
 \mathchardef\PLambda="7103
 \mathchardef\PSigma="7106
 \mathchardef\POmega="710A
 \mathchardef\PUpsilon="7107
                  
 \def\PB      {\ensuremath{B}\xspace}                 
                  
 \def\PD      {\ensuremath{D}\xspace}

 \def\PJ      {\ensuremath{J}\xspace}                 
 \def\PK      {\ensuremath{K}\xspace}

 \def\Pb      {\ensuremath{b}\xspace}                 
 \def\Pc      {\ensuremath{c}\xspace}

 \def\Pi      {\ensuremath{i}\xspace}

 \def\Ps      {\ensuremath{s}\xspace}

}

\def\mup        {\ensuremath{\Pmu^+}\xspace}
\def\mun        {\ensuremath{\Pmu^-}\xspace} % muon negative (\mum is taken)

\def\squark    {\ensuremath{\Ps}\xspace}

\def\cquark    {\ensuremath{\Pc}\xspace}

\def\bquark    {\ensuremath{\Pb}\xspace}

\def\pion  {\ensuremath{\Ppi}\xspace}

\def\pip   {\ensuremath{\pion^+}\xspace}
\def\pim   {\ensuremath{\pion^-}\xspace}

\def\kaon  {\ensuremath{\PK}\xspace}
  \def\Kbar  {\kern 0.2em\overline{\kern -0.2em \PK}{}\xspace}

\def\Kz    {\ensuremath{\kaon^0}\xspace}
\def\Kzb   {\ensuremath{\Kbar^0}\xspace}
\def\KzKzb {\ensuremath{\Kz \kern -0.16em \Kzb}\xspace}
\def\Kp    {\ensuremath{\kaon^+}\xspace}
\def\Km    {\ensuremath{\kaon^-}\xspace}

\def\KpKm  {\ensuremath{\Kp \kern -0.16em \Km}\xspace}

  \def\Dbar    {\kern 0.2em\overline{\kern -0.2em \PD}{}\xspace}
\def\D       {\ensuremath{\PD}\xspace}

\def\Dz      {\ensuremath{\D^0}\xspace}
\def\Dzb     {\ensuremath{\Dbar^0}\xspace}
\def\DzDzb   {\ensuremath{\Dz {\kern -0.16em \Dzb}}\xspace}
\def\Dp      {\ensuremath{\D^+}\xspace}
\def\Dm      {\ensuremath{\D^-}\xspace}

\def\DpDm    {\ensuremath{\Dp {\kern -0.16em \Dm}}\xspace}

\def\B       {\ensuremath{\PB}\xspace}
  \def\Bbar    {\kern 0.18em\overline{\kern -0.18em \PB}{}\xspace}

\def\Bu      {\ensuremath{\B^+}\xspace}

\def\Bp      {\ensuremath{\Bu}\xspace}

\def\Bs      {\ensuremath{\B^0_\squark}\xspace}

\def\jpsi     {\ensuremath{{\PJ\mskip -3mu/\mskip -2mu\Ppsi\mskip 2mu}}\xspace}

  \def\Y#1S{\ensuremath{\PUpsilon{(#1S)}}\xspace}% no space before {...}!

\def\Lbar {\ensuremath{\kern 0.1em\overline{\kern -0.1em\PLambda}}\xspace}

\def\to                 {\ensuremath{\rightarrow}\xspace}

\def\CP                {\ensuremath{C\!P}\xspace}

\def\AT#1     {\ensuremath{A_{\mathrm{T}}^{#1}}\xspace}           % 2

\def\C#1      {\ensuremath{\mathcal{C}_{#1}}\xspace}                       % 9
\def\Cp#1     {\ensuremath{\mathcal{C}_{#1}^{'}}\xspace}                    % 7
\def\Ceff#1   {\ensuremath{\mathcal{C}_{#1}^{\mathrm{(eff)}}}\xspace}        % 9  
\def\Cpeff#1  {\ensuremath{\mathcal{C}_{#1}^{'\mathrm{(eff)}}}\xspace}       % 7
\def\Ope#1    {\ensuremath{\mathcal{O}_{#1}}\xspace}                       % 2
\def\Opep#1   {\ensuremath{\mathcal{O}_{#1}^{'}}\xspace}                    % 7

\newcommand{\tev}{\ensuremath{\mathrm{\,Te\kern -0.1em V}}\xspace}
\newcommand{\gev}{\ensuremath{\mathrm{\,Ge\kern -0.1em V}}\xspace}
\newcommand{\mev}{\ensuremath{\mathrm{\,Me\kern -0.1em V}}\xspace}
\newcommand{\kev}{\ensuremath{\mathrm{\,ke\kern -0.1em V}}\xspace}
\newcommand{\ev}{\ensuremath{\mathrm{\,e\kern -0.1em V}}\xspace}
\newcommand{\gevc}{\ensuremath{{\mathrm{\,Ge\kern -0.1em V\!/}c}}\xspace}
\newcommand{\mevc}{\ensuremath{{\mathrm{\,Me\kern -0.1em V\!/}c}}\xspace}
\newcommand{\gevcc}{\ensuremath{{\mathrm{\,Ge\kern -0.1em V\!/}c^2}}\xspace}
\newcommand{\gevgevcccc}{\ensuremath{{\mathrm{\,Ge\kern -0.1em V^2\!/}c^4}}\xspace}
\newcommand{\mevcc}{\ensuremath{{\mathrm{\,Me\kern -0.1em V\!/}c^2}}\xspace}

\def\mum  {\ensuremath{\,\upmu\rm m}\xspace}

\def\invfb   {\ensuremath{\mbox{\,fb}^{-1}}\xspace}

\def\ps   {\ensuremath{{\rm \,ps}}\xspace}
\def\fs   {\ensuremath{\rm \,fs}\xspace}

\newcommand{\chisq}{\ensuremath{\chi^2}\xspace}

\def\gsim{{~\raise.15em\hbox{$>$}\kern-.85em
          \lower.35em\hbox{$\sim$}~}\xspace}
\def\lsim{{~\raise.15em\hbox{$<$}\kern-.85em
          \lower.35em\hbox{$\sim$}~}\xspace}

\def\sqs   {\ensuremath{\protect\sqrt{s}}\xspace}

\def\evtgen     {\mbox{\textsc{EvtGen}}\xspace}
\def\pythia     {\mbox{\textsc{Pythia}}\xspace}

\def\geant      {\mbox{\textsc{Geant4}}\xspace}

\def\tell1  {TELL1\xspace}
\def\ukl1   {UKL1\xspace}

\usepackage{cite} % Allows for ranges in citations
\usepackage{mciteplus}

\def\Bus     {\ensuremath{\B^{*+}}\xspace}

\def\Bus     {\ensuremath{\B^{*+}}\xspace}
\def\Bsone   {\ensuremath{\B_{\squark1}}\xspace}
\def\Bstwo   {\ensuremath{\B^{*}_{\squark2}}\xspace}

\begin{document}

%%%%%%%%%%%%%%%%%%%%%%%%%
%%%%% Title     %%%%%%%%%
%%%%%%%%%%%%%%%%%%%%%%%%%
\renewcommand{\thefootnote}{\fnsymbol{footnote}}
\setcounter{footnote}{1}

% %%%%%%% CHOOSE TITLE PAGE--------
%\onecolumn
% \input{title-LHCb-ANA}
% \input{title-LHCb-CONF}
% $Id: title-LHCb-PAPER.tex 21910 2012-06-25 21:26:22Z uegede $
% ===============================================================================
% Purpose: LHCb-PAPER journal paper title page template
% Author: 
% Created on: 2010-09-25
% ===============================================================================

%%%%%%%%%%%%%%%%%%%%%%%%%
%%%%%  TITLE PAGE  %%%%%%
%%%%%%%%%%%%%%%%%%%%%%%%%
\begin{titlepage}
\pagenumbering{roman}

% Header ---------------------------------------------------
\vspace*{-1.5cm}
\centerline{\large EUROPEAN ORGANIZATION FOR NUCLEAR RESEARCH (CERN)}
\vspace*{1.5cm}
\hspace*{-0.5cm}
\begin{tabular*}{\linewidth}{lc@{\extracolsep{\fill}}r}
\ifthenelse{\boolean{pdflatex}}% Logo format choice
{\vspace*{-2.7cm}\mbox{\!\!\!\includegraphics[width=.14\textwidth]{lhcb-logo.pdf}} & &}%
{\vspace*{-1.2cm}\mbox{\!\!\!\includegraphics[width=.12\textwidth]{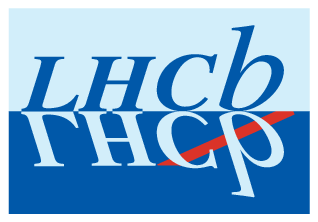}} & &}%
\\
 & & CERN-PH-EP-2012-340 \\  % ID 
 & & LHCb-PAPER-2012-030 \\  % ID 
 & & 26 November 2012 \\ % Date - Can also hardwire e.g.: 23 March 2010
 & & \\
\end{tabular*}

\vspace*{3.5cm}

% Title --------------------------------------------------
{\bf\boldmath\huge
\begin{center}
First observation of the decay $B_{s2}^*(5840)^0 \to B^{*+} K^-$ and studies of excited \Bs mesons
\end{center}
}

\vspace*{1.5cm}

% Authors -------------------------------------------------
\begin{center}
The LHCb collaboration\footnote{Authors are listed on the following pages.}
\end{center}

\vspace{\fill}

% Abstract -----------------------------------------------
\begin{abstract}
\noindent 
Properties of the orbitally excited ($L=1$) \Bs states are studied using 1.0\invfb of $p p$ collisions
at $\sqs =7$\tev collected with the \lhcb detector.
The first observation of the $B_{s2}^*(5840)^0$ meson decaying to $\Bus \Km$ is reported, and the corresponding branching fraction measured relative to the $\Bu \Km$ decay mode.
The \mbox{$\B_{s1}(5830)^0 \to \Bus \Km$} decay is observed as well. 
The width of the $B_{s2}^*(5840)^0$ state is measured for the first time and the masses of the
 two states are determined with the highest precision to date. 
The observation of the $B_{s2}^*(5840)^0 \to \Bus \Km$ decay favours the spin-parity assignment $J^P = 2^+$ for the $B_{s2}^*(5840)^0$ meson. 
In addition, the most precise measurement of the mass difference $m(\Bus)-m(\Bu) = 45.01 \pm 0.30 \, ({\rm stat}) \pm 0.23 \, ({\rm syst}) \mevcc$ is obtained.
\end{abstract}

\vspace*{1.5cm}

\begin{center}
Submitted to Phys. Rev. Lett.
\end{center}

\vspace{\fill}

\end{titlepage}

%%%%%%%%%%%%%%%%%%%%%%%%%%%%%%%%
%%%%%  EOD OF TITLE PAGE  %%%%%%
%%%%%%%%%%%%%%%%%%%%%%%%%%%%%%%%

%  empty page follows the title page ----
\newpage
\setcounter{page}{2}
\mbox{~}
\newpage

% Author List ----------------------------
%%%%%%%%%%%%%%%%%%%%%%%%%%%%%%%%%%%%%%%%%%
\centerline{\large\bf LHCb collaboration}
\begin{flushleft}
\small
R.~Aaij$^{38}$, 
C.~Abellan~Beteta$^{33,n}$, 
A.~Adametz$^{11}$, 
B.~Adeva$^{34}$, 
M.~Adinolfi$^{43}$, 
C.~Adrover$^{6}$, 
A.~Affolder$^{49}$, 
Z.~Ajaltouni$^{5}$, 
J.~Albrecht$^{35}$, 
F.~Alessio$^{35}$
% ...

{\footnotesize \it
% ...
$ ^{5}$Clermont Universit\'{e}, Universit\'{e} Blaise Pascal, CNRS/IN2P3, LPC, Clermont-Ferrand, France\\
$ ^{6}$CPPM, Aix-Marseille Universit\'{e}, CNRS/IN2P3, Marseille, France\\
% ...
$ ^{11}$Physikalisches Institut, Ruprecht-Karls-Universit\"{a}t Heidelberg, Heidelberg, Germany\\
% ...
$ ^{33}$Universitat de Barcelona, Barcelona, Spain\\
$ ^{34}$Universidad de Santiago de Compostela, Santiago de Compostela, Spain\\
$ ^{35}$European Organization for Nuclear Research (CERN), Geneva, Switzerland\\
% ...
$ ^{38}$Nikhef National Institute for Subatomic Physics, Amsterdam, The Netherlands\\
% ...
$ ^{43}$H.H. Wills Physics Laboratory, University of Bristol, Bristol, United Kingdom\\
% ...
$ ^{49}$Oliver Lodge Laboratory, University of Liverpool, Liverpool, United Kingdom\\
% ...
$ ^{n}$LIFAELS, La Salle, Universitat Ramon Llull, Barcelona, Spain\\
% ...
}
\end{flushleft}
%%%%%%%%%%%%%%%%%%%%%%%%%%%%%%%%%%%%%%%%%%

\cleardoublepage

%\twocolumn
% %%%%%%%%%%%%% ---------

\renewcommand{\thefootnote}{\arabic{footnote}}
\setcounter{footnote}{0}

%%%%%%%%%%%%%%%%%%%%%%%%%%%%%%%%
%%%%%  Table of Content   %%%%%%
%%%%%%%%%%%%%%%%%%%%%%%%%%%%%%%%
%%%% Uncomment next 2 lines if desired
%\tableofcontents
%\cleardoublepage

%%%%%%%%%%%%%%%%%%%%%%%%%
%%%%% Main text %%%%%%%%%
%%%%%%%%%%%%%%%%%%%%%%%%%

\pagestyle{plain} % restore page numbers for the main text
\setcounter{page}{1}
\pagenumbering{arabic}

% %%%%%%% CHOOSE --------
%% ----------------------------------
%% Line numbering on the left margin 
%% ----------------------------------
%% Uncomment during review phase. 
%% Comment it out before a final submission.
%%\linenumbers
%% --------------------------------
% %%%%%%%%%%%%% ---------

% You can include short sections directly in the main tex file.
% However, for larger papers it is desirable to split the text into
% several semiautonomous files, which can be revised independently.
% This is especially useful when developing a document in
% collaboration with several people, since then different parts can be
% edited independently.  This type of file organization is shown here.
% 

%% %%%%%%%%%%%%%%%%%%
%%  Page formatting
%% %%%%%%%%%%%%%%%%%%

%\usepackage{hyperref}
%\usepackage[all]{hypcap}

\noindent Heavy quark effective theory (HQET) describes mesons with one heavy and one light quark where the heavy quark is assumed to have infinite mass~\cite{Mannel:1996cn}.
It is an important tool for calculating
meson properties which may be modified by
physics beyond the Standard Model, such as \CP violation in charm meson decays~\cite{Bobrowski:2010xg} or the mixing and lifetimes of \B mesons~\cite{Lenz:2012mb}.
It also predicts the properties of excited \B and \Bs mesons~\cite{DiPierro:2001uu, Eichten:1993ub, Falk:1995th, Koponen:2007nr}, and precise measurements of
these properties are a sensitive test of the validity of the theory. 
Within HQET the \Bs mesons are characterised by three quantum numbers: the relative orbital angular momentum $L$ of the two quarks, the total angular
momentum of the light quark $j_q = |L \pm  \frac{1}{2}|$, and the total angular momentum of the \Bs meson $J = |j_q \pm \frac{1}{2}|$. 
For $L = 1$ there are four different possible ($J$, $j_q$) combinations, all with even parity. 
These are collectively termed the orbitally excited states.
Such
states 
can decay to $\Bp\Km$ and/or $\Bus\Km$ (the inclusion of charge-conjugate states is implied throughout this Letter), depending on their quantum numbers and mass values. The two states with $j_q = 1/2$, named $B_{s0}^*$ and $\Bsone^{\prime}$,
are expected to decay through an S-wave transition and to have a large $\mathcal O(100\mevcc)$ decay width. In contrast, the two states
with $j_q = 3/2$, named $B_{s1}(5830)^0$ and $B_{s2}^*(5840)^0$ (henceforth $\Bsone$ and $\Bstwo$ for brevity), are expected to decay through a D-wave transition and to have a narrow $\mathcal O(1\mevcc)$ decay width. Table~\ref{tab:bsstst} gives an overview of these states.

\begin{table}[hbt]
\begin{center}
\label{tab:bsstst}
\caption{Summary of the orbitally excited ($L=1$) \Bs states.}
\begin{tabular}{cccccc}
\hline 
 & $j_q$ & $J^P$ & \multicolumn{2}{c}{Allowed decay mode} & Mass (\mevcc)~\cite{PDG}\\ 
&  &  & $\Bu\Km$ & $\Bus \Km$ & \\
\hline
$B_{s0}^*$ & 1/2 & $0^+$ & yes & no & Unobserved\\
$\Bsone^{\prime}$ & 1/2 & $1^+$ & no & yes & Unobserved\\
$\Bsone$ & 3/2 & $1^+$ & no & yes & $5829.4 \pm 0.7$\\
$\Bstwo$ & 3/2 & $2^+$ & yes & yes & $5839.7 \pm 0.6$\\
\hline 
\end{tabular}
\end{center}
\end{table}

In this Letter a $1.0$~fb$^{-1}$ sample of data collected by the \lhcb detector is used to search
for the orbitally excited \Bs mesons in the mass distribution of $\Bu\Km$ pairs, where the \Bp mesons are selected in the four decay modes:
$\Bu\to J/\psi(\mup \mun) \Kp$, $\Bu\to \Dzb(\Kp\pim) \pip$, $\Bu\to \Dzb(\Kp\pim\pip\pim) \pip$, and $\Bu\to \Dzb(\Kp\pim) \pip\pim\pip$.
Two narrow peaks were observed in the $\Bp\Km$ mass distribution by the \cdf collaboration~\cite{Aaltonen:2007ah}. Putatively, they are identified with the states
of the $j_q=3/2$ doublet expected in HQET\cite{DiPierro:2001uu} and are named $\Bsone$ and $\Bstwo$. As the $\Bsone \to \Bu\Km$
decay is forbidden, one of the mass peaks observed is interpreted as the $\Bsone \to \Bus\Km$ decay followed by $\Bus \to \Bu \gamma$,  
where the photon is not observed.
This peak is shifted by the  $\Bus - \Bu$ mass
difference due to the missing momentum of the photon in the $\Bus \to \Bu \gamma$ decay. 
While the $\Bstwo \to \Bu\Km$ decay has been observed by the \dzero collaboration as well~\cite{Abazov:2007af}, a confirmation of the $\Bsone$ meson
is still missing. The identification of the $\Bsone$
  and $\Bstwo$ mesons in the $\Bp\Km$ mass spectrum is based on the
  expected mass splitting between the $j_q = 3/2$ states. The $\Bsone$
  and $\Bstwo$ widths are very sensitive to their masses, due to their
  proximity to the $BK$ and $B^*K$ thresholds. Measurements of the
  widths thus provide fundamental information concerning the nature of
  these states. In addition the $\Bsone$ and $\Bstwo$ quantum numbers
  have not yet been directly determined and the observation of other
  decay modes can constrain the spin-parity combinations of the states. In particular the
$\Bstwo \to \Bus \Km$ decay  has not yet been observed
but could manifest itself in the $\Bu\Km$ mass spectrum in a similar fashion to the corresponding \Bsone meson decay. The $\Bstwo \to \Bus\Km$ branching fraction relative
to $\Bstwo \to \Bp\Km$ is predicted to be between 2$\%$ and 10$\%$, depending on the $\Bstwo$ mass~\cite{Godfrey:1986wj, Colangelo:2012xi, Zhong:2008kd, Wang:2012pf}. 

Recently the Belle collaboration has reported observation of charged
bottomonium-like $Z_b(10610)^+$ and $Z_b(10650)^+$
  states~\cite{Belle:2011aa, Adachi:2012cx}, that could be interpreted as $\B\Bbar^{*}$ and $\B^{*}\Bbar^{*}$ 
molecules respectively~\cite{Bondar:2011ev}. To test this interpretation, improved measurements of the \Bus mass are necessary, and can be obtained from                                         
the difference in peak positions between $\Bstwo \to \Bus\Km$ and
$\Bstwo \to \Bu\Km$ decays in the $\Bp\Km$ mass
  spectrum. 

The \lhcb detector~\cite{Alves:2008zz} is a single-arm forward
spectrometer covering the pseudo-rapidity range $2<\eta <5$, designed
for studying particles containing \bquark or \cquark quarks. 
The detector includes a high-precision tracking system consisting of a
silicon-strip vertex detector surrounding the $pp$ interaction region,
a large-area silicon-strip detector located upstream of a dipole
magnet with a bending power of about $4{\rm\,Tm}$, and three stations
of silicon-strip detectors and straw drift tubes placed
downstream. The combined tracking system has a momentum resolution
($\Delta p/p$), that varies from 0.4\% at 5\gevc to 0.6\% at 100\gevc, and a decay time resolution of 50\fs.
The resolution of the impact parameter, the transverse distance of closest approach between the track and a primary interaction, is about 20\mum for tracks with large
transverse momentum.  The transverse component is measured in the plane normal to the beam axis.
Charged hadrons are identified using two 
ring-imaging Cherenkov detectors. Photon, electron and hadron
candidates are identified by a calorimeter system consisting of
scintillating-pad and pre-shower detectors, an electromagnetic
calorimeter, and a hadronic calorimeter. Muons are identified by a
system composed of alternating layers of iron and multiwire
proportional chambers. 

The trigger system~\cite{Aaij:2012me} consists of a hardware stage, based on information from the calorimeter and muon systems, 
followed by a software stage that applies a full event reconstruction.
Events likely to contain a \B meson are selected by searching for a dimuon vertex detached from the primary interaction
or two-, three-, and four-track vertices detached from the primary interaction which have high total
transverse momentum. These are respectively referred to as dimuon and topological triggers.

The samples of simulated events used in this analysis
are based on the \pythia 6.4 generator \cite{Sjostrand:2006za}, with a choice of parameters
specifically configured for LHCb \cite{Belyaev:2011zza}. The \evtgen package \cite{Lange:2001uf}
describes the decay of the \B mesons, and the \geant toolkit \cite{Allison:2006ve, *Agostinelli:2002hh, LHCb-PROC-2011-006} is used to simulate the
detector response. QED radiative corrections are generated with the  {\mbox{\textsc{Photos}}\xspace} package~\cite{Golonka:2005pn}.

\begin{figure*}[t]
 \centering
\includegraphics[width=.38\textwidth]{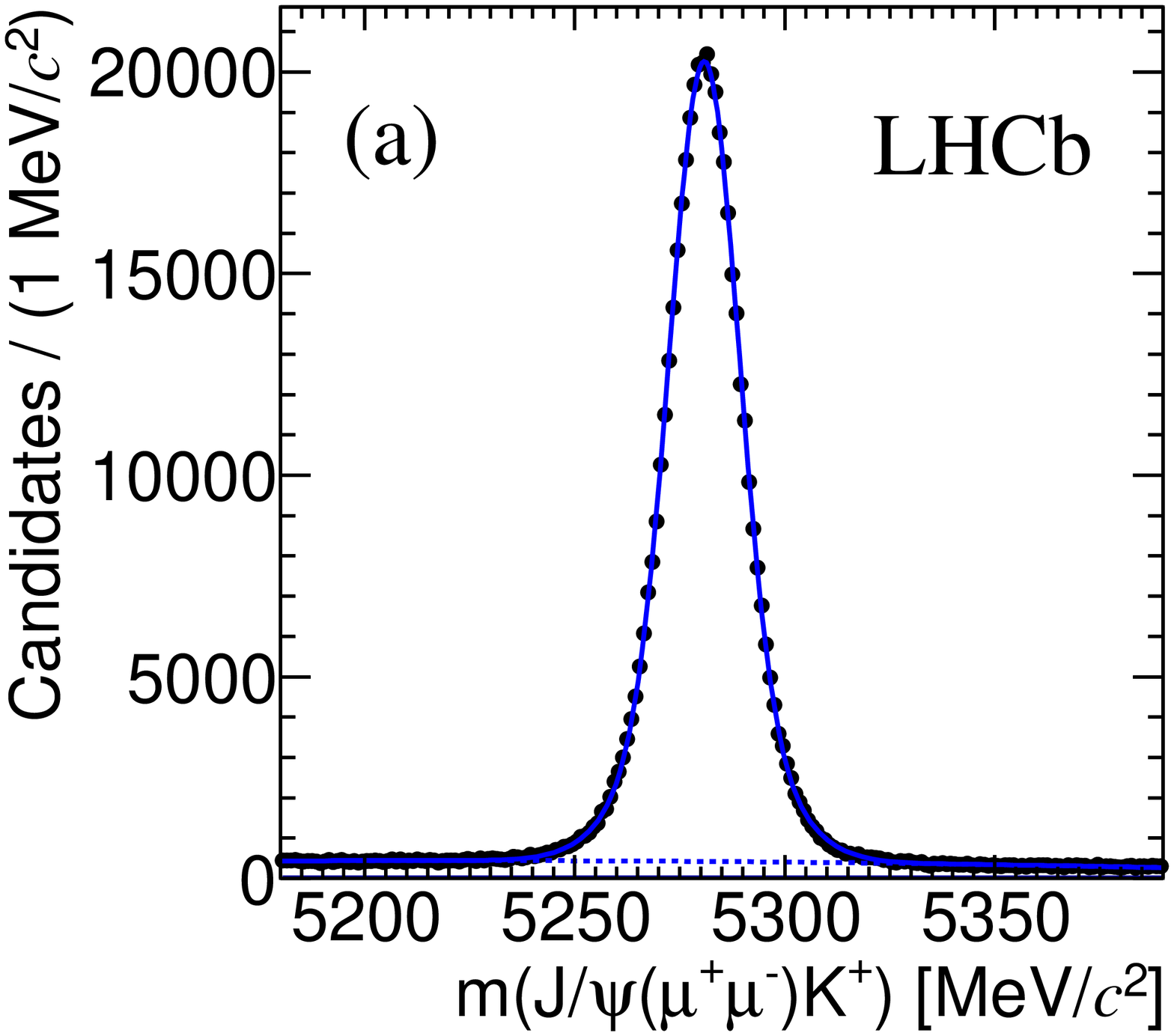}
\includegraphics[width=.38\textwidth]{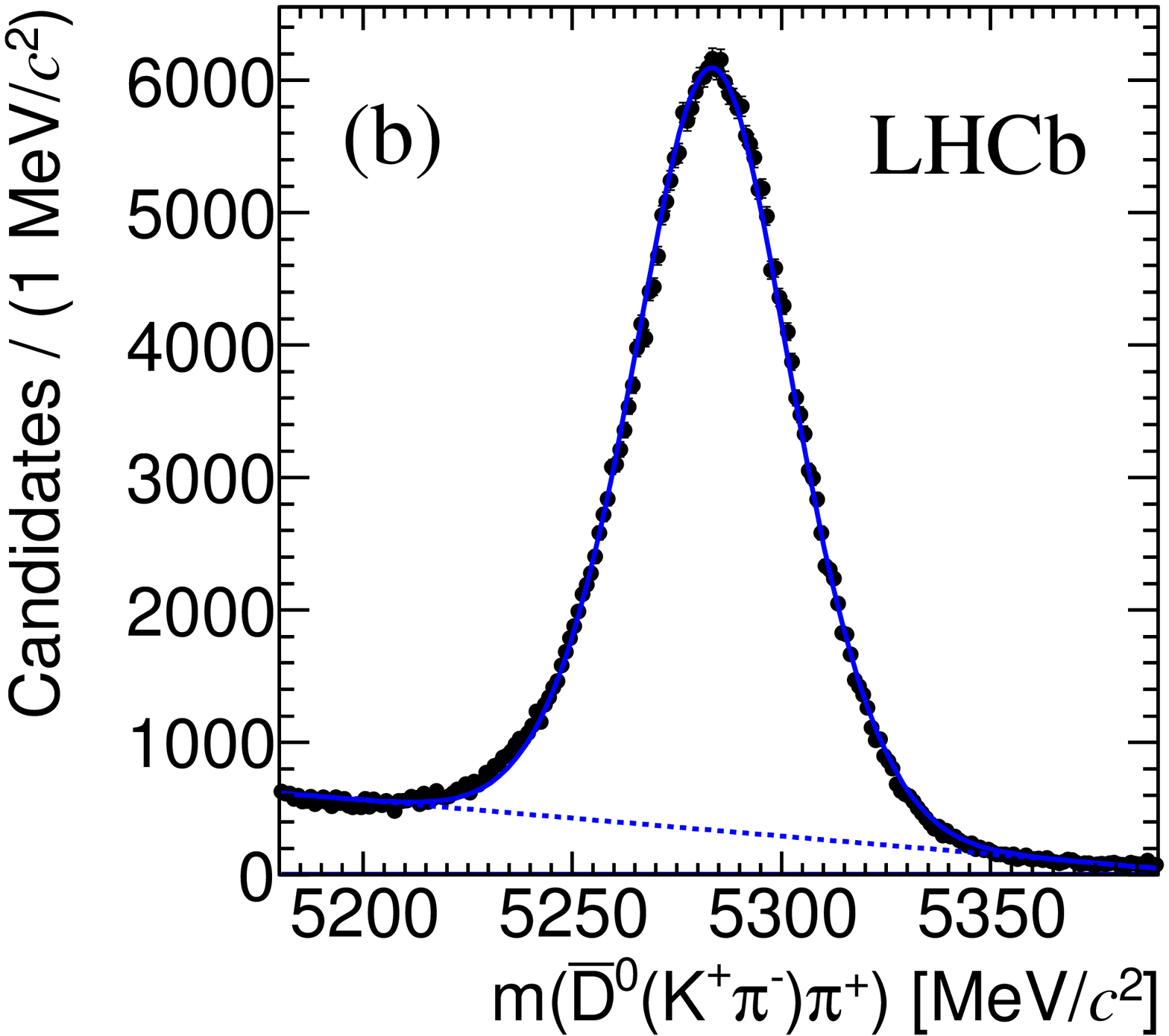}\\ \vspace{0.5cm}
\includegraphics[width=.38\textwidth]{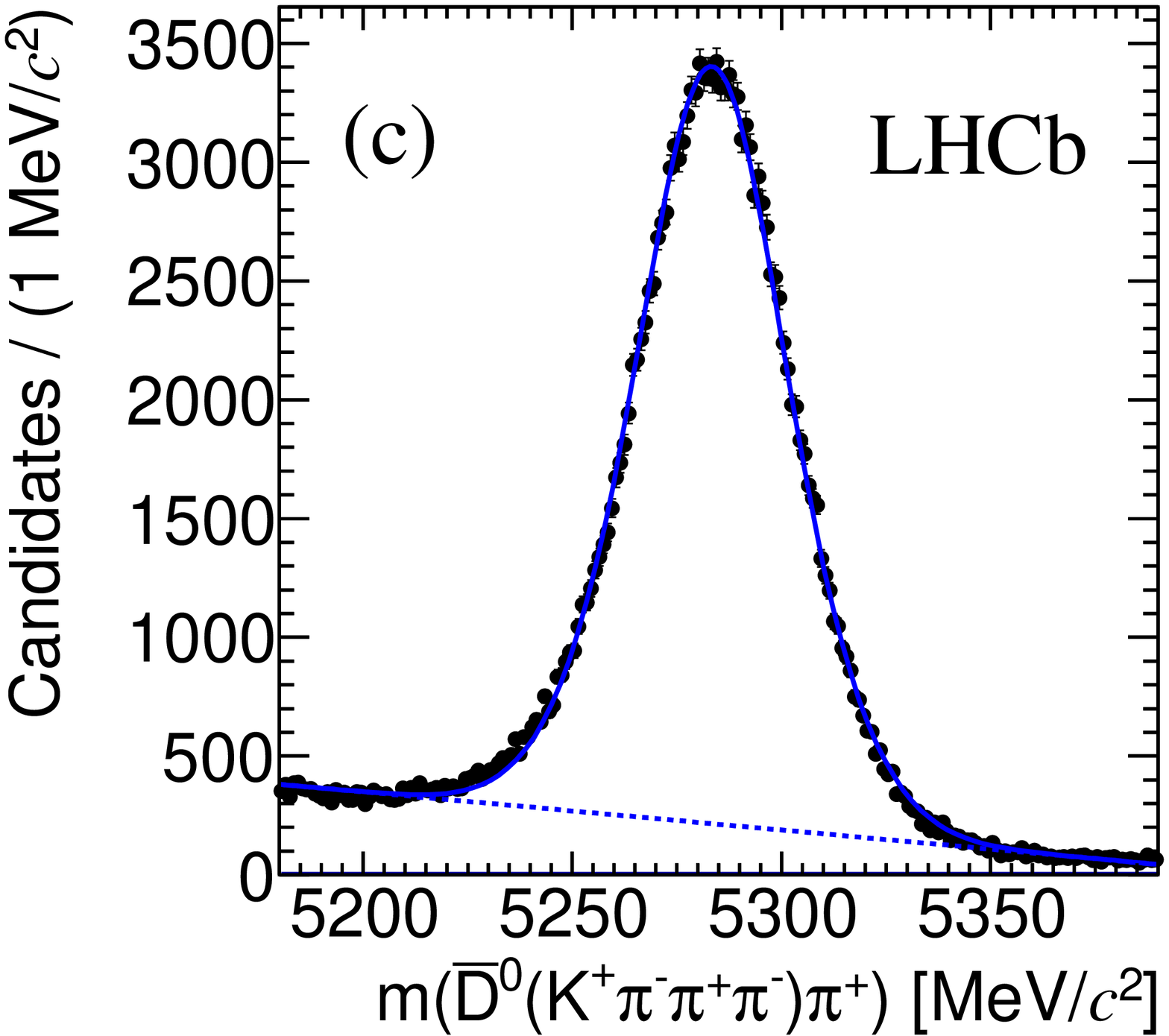}
\includegraphics[width=.38\textwidth]{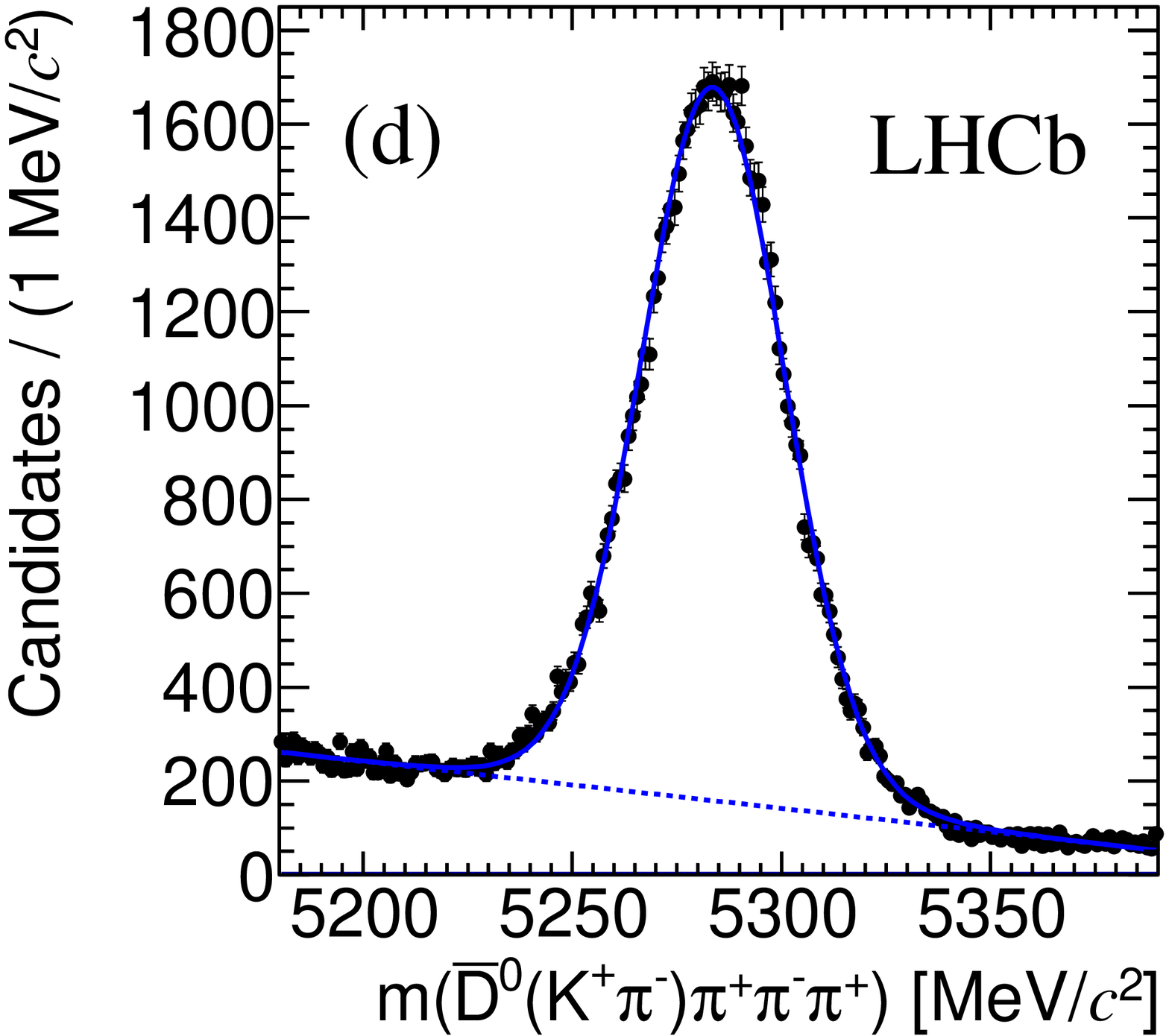}
\caption{
Invariant mass spectra of the final \Bu candidates.
The signal lineshape is fitted with a double Gaussian distribution, while the background
is modelled with a second order polynomial .
(a) $\Bu\to J/\psi \Kp$, (b) $\Bu\to \Dzb(\Kp\pim) \pip$, (c) $\Bu\to \Dzb(\Kp\pim\pip\pim) \pip$, and (d) $\Bu\to \Dzb(\Kp\pim) \pip\pim\pip$ decays. 
The \jpsi and \Dz masses are constrained to their world average values.}
 \label{fig:bsignals}
\end{figure*}

In the offline analysis the \B mesons are reconstructed 
using a set of loose selection criteria to suppress the majority of the combinatorial backgrounds. 
The $\Bu\to J/\psi \Kp$ selection requires a \Bu candidate with a transverse momentum
of at least 2\gevc and a decay time of at least 0.3\ps.
For the other decay modes, the selection explicitly
requires that the topological trigger, which selected the event, is based exclusively on tracks from which the \B meson candidate is formed.
Additional loose selection requirements are placed on variables related to the \B meson production and decay such as
transverse momentum and quality of the track fits for the decay products,
detachment of the \Bu
candidate from the primary interaction, whether the momentum of the \Bu candidate points back to the primary interaction, and the impact parameter \chisq.
The impact parameter \chisq is defined as the difference between the \chisq of the primary vertex reconstructed with and without the considered track.

Following these selections, \Bu signals are visible above backgrounds in all four decay modes. In order to improve their purity,
four boosted decision tree classifiers~\cite{Breiman,*Roe} are trained on variables common to all four decay modes: the transverse momenta and impact parameters
of the final state tracks, the transverse momentum and impact parameter of the \Bu candidate, the detachment of the \Bu candidate from the primary
interaction, 
the cosine of the angle between the \Bu candidate momentum and the direction of flight from the primary vertex to the decay vertex, 
the fit \chisq of the tracks,
and particle identification information. 
The classifier is trained on data using the {\it sWeights}
technique~\cite{Pivk:2004ty}, with the \Bu candidate mass as a  discriminating variable, to unfold the signal and background distributions. The cut on the classifier response
is chosen by optimizing the significance of each \Bu signal. The final mass distributions for the \Bu candidates are shown in Fig.~\ref{fig:bsignals}. 

The \Bu candidate mass spectra are fitted using a double Gaussian
function for the signal and a second order polynomial for the
background. The average mass resolution, $\sigma_{\Bu}$, is defined as
the weighted average of the Gaussian widths. The
purities of the samples, defined as the fraction of the signal events in a
$\pm 2 \sigma_{\Bu}$ mass region, are $96\%$, $91\%$, $90\%$, and  $85\%$ for
the $\Bu\to J/\psi \Kp$, $\Bu\to \Dzb(\Kp\pim) \pip$, 
$\Bu\to \Dzb(\Kp\pim\pip\pim) \pip$, and $\Bu\to \Dzb(\Kp\pim) \pip\pim\pip$ decays  respectively.
The \Bu candidates, within a $\pm 2 \sigma_{\Bu}$ mass region,
are selected for each decay mode. A sample of about $1\,000\,000$ \Bu candidates is obtained and combined  with any track of opposite charge that is
identified as a kaon.

Multiple $p p$ interactions can occur in \lhc bunch crossings. In order to reduce combinatorial backgrounds, the \Bu and kaon candidates are
required to be consistent with coming from the same interaction point. The signal purity is improved by a boosted decision tree classifier, whose inputs are
the \Bu and the kaon transverse momenta, the log-likelihood difference between the kaon and pion hypotheses, and the vertex fit and impact parameter \chisq.
The training is performed using simulated events for the signal and the like-charge $\Bu\Kp$ candidates in the data for the background. 
The same selection is subsequently applied to all \Bu decay modes. The cut on the classifier response is chosen by optimizing the significance of the $\Bstwo \to \Bu \Km$ signal. It retains 
57\% of the signal events and rejects 92\% of the background events.
In order to improve the mass resolution, the $\Bu\Km$ mass fits are performed constraining the $J/\psi$ (or $\Dz$) and \Bu
particles to their respective world average masses~\cite{PDG} and constraining the \Bu and \Km momenta to point to the associated primary vertex.

\begin{figure}[t]
 \centering
\includegraphics[width=.95\textwidth]{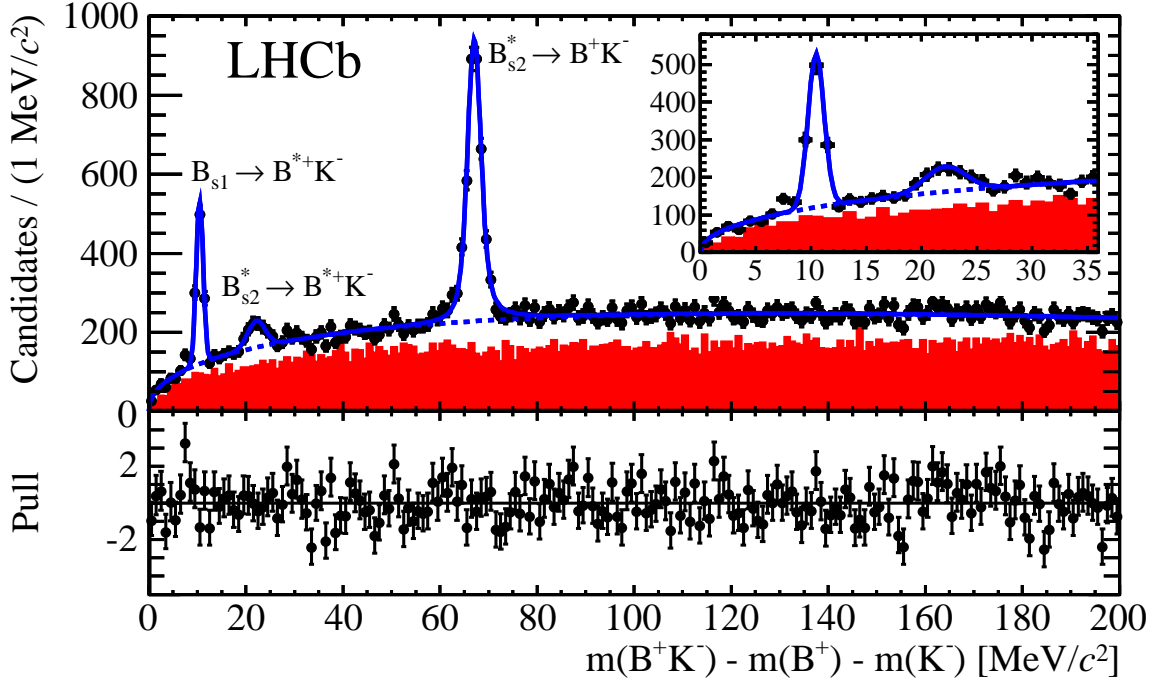}
\caption{Mass difference distribution $m(\Bu\Km)-m(\Bu)-m(\Km)$. The three peaks are identified as (left) $\Bsone\to \Bus \Km$, (middle)
$\Bstwo\to \Bus \Km$, and (right) $\Bstwo\to \Bu \Km$. The total fit function is shown as a solid blue line, while
the shaded red region is the spectrum of like-charge $\Bu\Kp$ combinations. The inset shows an expanded view of the $\Bsone/\Bstwo \to \Bus \Km$ signals. The bottom plot shows the fit pulls. }
 \label{fig:bsssignals}
\end{figure}

Figure~\ref{fig:bsssignals} shows the mass difference for the selected candidates, summed over all \Bu decay modes.
The mass difference is defined as $Q \equiv m(\Bu\Km)-m(\Bu)-m(\Km)$ where  $m(\Bu)$ and $m(\Km)$ are  the known masses of the \Bu and \Km mesons~\cite{PDG}, respectively.
The two narrow peaks at 10 and $67\mevcc$ are identified as the $\Bsone\to \Bus \Km$ and \mbox{$\Bstwo\to \Bu \Km$} signals, respectively, as previously observed.
In addition, a smaller structure is seen around
$20\mevcc$, identified as the previously unobserved
$\Bstwo\to \Bus \Km$ decay mode.

Simulated events are used to compute the detector
  resolutions corresponding to the three signals. The values obtained
  are increased by 20$\%$ to account for differences between the \Bu
  resolutions in data and simulated events. The corrected resolutions
  are 0.4~\mevcc, 0.6~\mevcc and 1.0~\mevcc for the $\Bsone\to \Bus \Km$,
  $\Bstwo\to \Bus \Km$, and $\Bstwo\to \Bu \Km$ signals
  respectively. A discrepancy of 40\% between the mass resolutions 
in  data and simulated events is observed for decays with small Q values,
such as $D^{*+} \to \Dz \pip$. Therefore we assign an uncertainty of
$\pm 20\%$ to the resolution in the systematic studies.

An unbinned fit of the mass difference distribution is performed to extract the $Q$ values and event yields of the three peaks.
The $\Bstwo \to \Bu\Km$ signal is parameterized by a relativistic Breit-Wigner function with natural width $\Gamma$
convolved with a Gaussian function that accounts for the detector resolution. 
Its width is fixed to the value obtained from simulated events.
The lineshapes of the \mbox{$\Bsone/\Bstwo \to \Bus \Km$} signals,
expected to be Breit-Wigner functions in the $\Bus\Km$ mass spectrum, are affected by the phase space and the angular distribution of the decays as the photon is not reconstructed.
The resulting shapes can not be properly simulated due to the lack of knowledge of the $\Bsone/\Bstwo$ properties. Therefore a  Gaussian function is used for each $\Bsone/\Bstwo \to \Bus \Km$ signals as  effective parameterization.
The background is modelled by a threshold function, $f(Q) = Q^\alpha e^{\beta Q + \delta}$, where $\alpha$, $\beta$ and $\delta$ are free parameters in the fit. Its analytical form is verified
by fitting the like charge $\Bu \Kp$ combinations where no signal is expected. 

The parameters allowed to vary in the fit are: the yield $N_{\Bstwo\to \Bu \Km}$,  the yield ratios
$N_{\Bsone\to \Bus \Km}/N_{\Bstwo\to \Bu \Km}$ and $N_{\Bstwo\to \Bus \Km}/N_{\Bstwo\to \Bu \Km}$,
 the $Q$ values of the \mbox{$\Bsone \to \Bus \Km$} and $\Bstwo\to \Bu \Km$ signals, the mass difference
between the $\Bstwo\to \Bu \Km$ and \mbox{$\Bstwo\to \Bus \Km$} peaks, the natural width of the \Bstwo state, the Gaussian widths of $\Bsone/\Bstwo \to \Bus \Km$ signals and the parameters of the threshold function.
From the yield ratios, the relative branching fraction
\begin{equation}
\frac{\mathcal B(\Bstwo \to \Bus \Km)}{\mathcal B(\Bstwo \to \Bu \Km)} = \frac{N_{\Bstwo\to \Bus \Km}}{N_{\Bstwo\to \Bu \Km}} \times \epsilon^{\rm rel}_{2,2} = R^{\Bstwo} 
\end{equation}
is measured. The \Bsone to \Bstwo ratio of production cross-sections times 
the ratio of branching fractions of $\Bsone\to \Bus \Km$ relative to that of $\Bstwo\to \Bu \Km$ is also determined from
\begin{equation}
\frac{\sigma(pp\to \Bsone X)\mathcal B (\Bsone \to \Bus \Km)}{\sigma(pp\to \Bstwo X)\mathcal B (\Bstwo \to \Bu \Km)} = \frac{N_{\Bsone\to \Bus \Km}}{N_{\Bstwo\to \Bu \Km}} \times \epsilon^{\rm rel}_{1,2} = \sigma^{\Bsone/\Bstwo}R^{\Bsone/\Bstwo}
\end{equation}
\noindent These ratios are corrected by the relative selection 
efficiencies, $\epsilon^{\rm rel}_{2,2} =1.05 \pm 0.02$ and
$\epsilon^{\rm rel}_{1,2}=1.03\pm0.01$, using simulated decays. 
The fit results are given in Table~\ref{t:fitresults}. 
The widths of
the two Gaussian functions are $0.73\pm0.04\mevcc$ and $1.9\pm0.3\mevcc$ for the $\Bsone \to
\Bus \Km$  and $\Bstwo \to \Bus \Km$  signals respectively.
A binned $\chi^2$ test 
gives a confidence level of $43\%$ for the fit.

\begin{table}
  \begin{center}
    \label{t:fitresults}
    \caption{Results of the fit to the mass difference distributions $m(\Bu\Km)-m(\Bu)-m(\Km)$. The first uncertainties are statistical and the second are systematic.}
  \begin{tabular}{@{}l@{}l @{\hspace{0.2cm}}r@{}c@{}l@{}l @{\hspace{0.25cm}}r@{}c@{}l@{}l@{}}
   \hline 

\multicolumn{2}{@{}l}{Parameter}&\multicolumn{4}{c}{Fit result}&\multicolumn{4}{c@{}}{Best previous measurement}\\
\hline
   $m(\Bsone)$&$ - \, m(\Bus) - m(\Km)$ & $10.46 \, \pm \,$ & $0.04$ & $\, \pm \, 0.04$&$\mevcc$&$10.73 \, \pm \,$ & $0.21$ & $\, \pm \, 0.14$&$\mevcc$~\cite{Aaltonen:2007ah} \\
   $m(\Bstwo)$&$- \, \, m(\Bu) \, - m(\Km)$                     & $67.06 \, \pm \,$ & $0.05$ & $\, \pm \, 0.11$&$\mevcc$&$66.96 \, \pm \,$ & $0.39$ & $\, \pm \, 0.14$&$\mevcc$~\cite{Aaltonen:2007ah}\\
   $m(\Bus)   $&$- \, \, m(\Bu)$                                         & $45.01 \, \pm  \,$ & $0.30$ & $\, \pm \, 0.23$&$\mevcc$&\multicolumn{3}{c}{$45.6 \pm 0.8$} & $\mevcc$~\cite{Wu:1991xn} \\
 \multicolumn{2}{@{}l}{$\Gamma(\Bstwo)$} & $1.56 \, \pm \,$ & $0.13$ &$\, \pm \, 0.47$&$\mevcc$\\
 \multicolumn{2}{@{}l}{{\large $\frac{\mathcal B(\Bstwo \to \Bus \Km)}{\mathcal B(\Bstwo \to \Bu \Km)}$} }& $(9.3\phantom{0} \, \pm \, $ & $1.3\phantom{0}$ & $\, \pm \, 1.2)$&\%\\
 \multicolumn{2}{@{}l}{{\large $\frac{\sigma(pp\to \Bsone X)\mathcal B (\Bsone \to \Bus \Km)}{\sigma(pp\to \Bstwo X)\mathcal B (\Bstwo \to \Bu \Km)}$}} & $(23.2\phantom{0} \, \pm \,$ & $1.4\phantom{0}$ & $\, \pm \, 1.3)$&\%\\
 \multicolumn{2}{@{}l}{$N_{\Bsone \to \Bus \Km}$} & \multicolumn{3}{c}{$750\pm 36$}\\
 \multicolumn{2}{@{}l}{$N_{\Bstwo \to \Bus \Km}$} & \multicolumn{3}{c}{$307\pm 46$}\\
 \multicolumn{2}{@{}l}{$N_{\Bstwo \to \Bp \Km}$} & \multicolumn{3}{c}{$3140\pm 100$}\\
\hline
  \end{tabular}
  \end{center}
\end{table}

To determine the significance of the $\Bstwo \to \Bus \Km$ signal, a similar maximum likelihood fit is performed, where
all parameters of the signal are fixed according to expectation, except its yield.
The likelihood of this fit is compared to the 
result of a fit where the yield of the signal is fixed to zero.
The statistical significance of the $\Bstwo \to \Bus \Km$ signal is $8 \sigma$.

A number of systematic uncertainties are considered. For the signal model, 
 the signal shape is changed to a double Gaussian function and an alternative threshold function is used for the background.
The changes in the fit results are assigned as the associated uncertainties.
The \Bu decay modes are fitted independently to test for effects that may be related to differences in their selection requirements.
For each observable quoted in Table~\ref{t:fitresults}, the difference between the weighted average of these independent
fits and the global fit is taken as a systematic uncertainty.
Additional systematic uncertainties are assigned based on the change in the results when varying
the selection criteria and the \Bu signal region.
The detector resolution of $\Bstwo \to \Bu \Km$ signal is varied by $\pm 20\%$.
In addition, the momentum scale in the processing of the data used in this analysis is varied within the estimated uncertainty of $0.15\%$.
The corresponding uncertainty on the measured masses is assigned
as a systematic uncertainty.
The uncertainty on the determination of the selection efficiency ratios caused by finite samples of simulated events
is taken as a systematic uncertainty for the branching fractions. Finally simulated events are used to estimate the mass shifts
of the $\Bsone/\Bstwo \to \Bus\Km$ signals from the nominal values when the radiated photon is 
excluded from their reconstructed decays.
The absolute systematic uncertainties are given in
Table~\ref{t:fitsyst}. 
The $\Bstwo \to \Bus \Km$ signal is observed with the
expected frequency in each of the four resconstructed decay modes and the systematic error for the
  $\frac{\mathcal B(\Bstwo \to \Bus \Km)}{\mathcal B(\Bstwo \to \Bu
    \Km)}$ branching fraction ratio, related to the different \Bu decay
  modes, is small.
The final results are shown in Table~\ref{t:fitresults}. 
The measured mass differences are more precise than the previous best
measurements of a factor two at least. The measured $\frac{\mathcal B(\Bstwo \to \Bus \Km)}{\mathcal B(\Bstwo \to \Bu \Km)}$ branching fraction ratio and \Bstwo width
are in good agreement with theoretical
predictions~\cite{Colangelo:2012xi, Zhong:2008kd, Wang:2012pf}.

The mass differences given in Table~\ref{t:fitresults} are translated into absolute masses by adding the masses of the \Bu and kaon~\cite{PDG} and,
in the case of the \Bsone meson, the $\Bus - \Bu$ mass difference measured in this Letter. The results are

$$\setlength\arraycolsep{0.1em}
\begin{array}{rccccccccl}
m(\Bus)     & =& 5324.26  &\pm& 0.30 &\pm & 0.23 & \pm & 0.17&\mevcc,\\
m(\Bsone) & = &5828.40 &\pm& 0.04& \pm& 0.04& \pm& 0.41& \mevcc,\\
m(\Bstwo) &=& 5839.99 &\pm& 0.05 &\pm& 0.11 &\pm& 0.17&\mevcc,
\end{array}
$$
where the first uncertainty is statistical and the second is systematic. The third uncertainty corresponds to the uncertainty on the \Bu mass~\cite{PDG} and, in the case of the \Bsone mass measurement, the uncertainty on the $\Bus - \Bu$ mass difference measured in this analysis.

\begin{table*}
  \begin{center}
    \label{t:fitsyst}
    \caption{Absolute systematic uncertainties for each measurement, which are assumed to be independent and are added in quadrature.}
\resizebox{0.99\textwidth}{!}{ 
  \begin{tabular}{ lcccccc }
    \hline
    Source                           & $Q(\Bsone)$ & $Q(\Bstwo)$ & $m(\Bus)-m(\Bu)$ & $\Gamma(\Bstwo)$ & $R^{\Bstwo}$ & $\sigma^{\Bsone/\Bstwo}R^{\Bsone/\Bstwo}$\\
                                             & $(\mevcc)$   & $(\mevcc)$  &   $(\mevcc)$            &   $(\mevcc)$           &       (\%)             & (\%)\\
    \hline
    Fit model                              & $0.00$   & $0.02$    &  $0.03$  & $0.01$&$0.2$ & $0.5$\\
    $\Bu$ decay mode                & $0.01$   & $0.01$   &  $0.02$  & $0.01$&$0.1$ & $0.1$ \\
    Selection                              & $0.03$    & $0.02$   &  $0.19$   & $0.05$&$1.1$ & $0.6$\\
    $\Bu$ signal region              & $0.01$   & $0.03$    &  $0.11$   & $0.07$&$0.2$ & $0.4$\\
    Mass resolution                 & $0.00$    & $0.01$   &  $0.02$   & $0.46$&$0.2$ & $0.9$\\
    Momentum scale                  & $0.02$   & $0.10$    & $0.03$  &           -&        - & -\\
   Efficiency  ratios               &     -         &   -          &     -         &       -    &$0.2$ & $0.2$\\
   Missing photon                   & $0.01$    & -             & $0.01$   &    -      &   -      &   -\\
    \hline
    Total                                     & $0.04$   & $0.11$   &  $0.23$      & $0.47$&$1.2$& $1.3$\\
    \hline 
  \end{tabular}
} 
  \end{center}
\end{table*}

The significance of the non-zero $\Bstwo$ width is determined by comparing the
likelihood for the nominal fit with a fit in which the width is fixed to
zero. To account for systematic effects, the minimum
  $\sqrt{2 \Delta \rm{log} \mathcal L}$ among all systematic variations is taken;
  the significance including systematic uncertainties is $9\sigma$.

In conclusion, using 1.0\invfb of data collected with the \lhcb detector at $\sqs =7$\tev, the decay mode $\Bstwo \to \Bus \Km$ is observed for the first time
and its branching fraction measured relative to that of $\Bstwo \to \Bu \Km$.  The observation of the \Bstwo meson decaying to two pseudoscalars ($\Bstwo \to \Bu \Km$) and to a vector and a pseudoscalar ($\Bstwo \to \Bus \Km$) favours the assignment of $J^P = 2^+$ for this state. The \Bstwo width is measured for the first time, while the masses of the \Bsone and \Bstwo states are measured with the highest
precision to date and are consistent with previous measurements~\cite{Aaltonen:2007ah, Abazov:2007af}.
Finally, the observed $\Bstwo \to \Bus \Km$ decay is used 
to make the most precise measurement to date of
the $\Bus-\Bu$ mass difference. This measurement, unlike others reported in the literature, does not require
the reconstruction of the soft photon from \Bus decays and therefore has significantly smaller systematic uncertainty.
High precision measurements of the \Bus mass are important for the
understanding of the exotic $Z_b^+$ states recently
observed\cite{Belle:2011aa}. Using the \Bus mass measured
  in this analysis, we compute that the $Z_b(10610)^+$ and
  $Z_b(10650)^+$ masses are $3.69 \pm 2.05 \mevcc$ and $3.68 \pm 1.71 \mevcc$ above the $\B\Bbar^{*}$ and
  $\B^{*}\Bbar^{*}$ thresholds respectively.

\section*{Acknowledgements}

\noindent We express our gratitude to our colleagues in the CERN
accelerator departments for the excellent performance of the LHC. We
thank the technical and administrative staff at the LHCb
institutes. We acknowledge support from CERN and from the national
agencies: CAPES, CNPq, FAPERJ and FINEP (Brazil); NSFC (China);
CNRS/IN2P3 and Region Auvergne (France); BMBF, DFG, HGF and MPG
(Germany); SFI (Ireland); INFN (Italy); FOM and NWO (The Netherlands);
SCSR (Poland); ANCS/IFA (Romania); MinES, Rosatom, RFBR and NRC
``Kurchatov Institute'' (Russia); MinECo, XuntaGal and GENCAT (Spain);
SNSF and SER (Switzerland); NAS Ukraine (Ukraine); STFC (United
Kingdom); NSF (USA). We also acknowledge the support received from the
ERC under FP7. The Tier1 computing centres are supported by IN2P3
(France), KIT and BMBF (Germany), INFN (Italy), NWO and SURF (The
Netherlands), PIC (Spain), GridPP (United Kingdom). We are thankful
for the computing resources put at our disposal by Yandex LLC
(Russia), as well as to the communities behind the multiple open
source software packages that we depend on.

\addcontentsline{toc}{section}{References}
\bibliographystyle{LHCb}
\bibliography{main}

\ifx\mcitethebibliography\mciteundefinedmacro
\PackageError{LHCb.bst}{mciteplus.sty has not been loaded}
{This bibstyle requires the use of the mciteplus package.}\fi
\providecommand{\href}[2]{#2}
\begin{mcitethebibliography}{10}
\mciteSetBstSublistMode{n}
\mciteSetBstMaxWidthForm{subitem}{\alph{mcitesubitemcount})}
\mciteSetBstSublistLabelBeginEnd{\mcitemaxwidthsubitemform\space}
{\relax}{\relax}

\bibitem{Mannel:1996cn}
T.~Mannel, \ifthenelse{\boolean{articletitles}}{{\it {Review of heavy quark
  effective theory}}, }{}\href{http://arxiv.org/abs/hep-ph/9611411}{{\tt
  arXiv:hep-ph/9611411}}\relax
\mciteBstWouldAddEndPuncttrue
\mciteSetBstMidEndSepPunct{\mcitedefaultmidpunct}
{\mcitedefaultendpunct}{\mcitedefaultseppunct}\relax
\EndOfBibitem
\bibitem{Bobrowski:2010xg}
M.~Bobrowski, A.~Lenz, J.~Riedl, and J.~Rohrwild,
  \ifthenelse{\boolean{articletitles}}{{\it {How large can the SM contribution
  to CP violation in $D^0-\bar D^0$ mixing be?}},
  }{}\href{http://dx.doi.org/10.1007/JHEP03(2010)009}{JHEP {\bf 03} (2010)
  009}, \href{http://arxiv.org/abs/1002.4794}{{\tt arXiv:1002.4794}}\relax
\mciteBstWouldAddEndPuncttrue
\mciteSetBstMidEndSepPunct{\mcitedefaultmidpunct}
{\mcitedefaultendpunct}{\mcitedefaultseppunct}\relax
\EndOfBibitem
\bibitem{Lenz:2012mb}
A.~Lenz, \ifthenelse{\boolean{articletitles}}{{\it {Theoretical update of
  $B$-mixing and lifetimes}}, }{}\href{http://arxiv.org/abs/1205.1444}{{\tt
  arXiv:1205.1444}}\relax
\mciteBstWouldAddEndPuncttrue
\mciteSetBstMidEndSepPunct{\mcitedefaultmidpunct}
{\mcitedefaultendpunct}{\mcitedefaultseppunct}\relax
\EndOfBibitem
\bibitem{DiPierro:2001uu}
M.~Di~Pierro and E.~Eichten, \ifthenelse{\boolean{articletitles}}{{\it {Excited
  heavy-light systems and hadronic transitions}},
  }{}\href{http://dx.doi.org/10.1103/PhysRevD.64.114004}{Phys.\ Rev.\  {\bf
  D64} (2001) 114004}, \href{http://arxiv.org/abs/hep-ph/0104208}{{\tt
  arXiv:hep-ph/0104208}}\relax
\mciteBstWouldAddEndPuncttrue
\mciteSetBstMidEndSepPunct{\mcitedefaultmidpunct}
{\mcitedefaultendpunct}{\mcitedefaultseppunct}\relax
\EndOfBibitem
\bibitem{Eichten:1993ub}
E.~J. Eichten, C.~T. Hill, and C.~Quigg,
  \ifthenelse{\boolean{articletitles}}{{\it {Properties of orbitally excited
  heavy-light ($Q\bar q$) mesons}},
  }{}\href{http://dx.doi.org/10.1103/PhysRevLett.71.4116}{Phys.\ Rev.\ Lett.\
  {\bf 71} (1993) 4116}, \href{http://arxiv.org/abs/hep-ph/9308337}{{\tt
  arXiv:hep-ph/9308337}}\relax
\mciteBstWouldAddEndPuncttrue
\mciteSetBstMidEndSepPunct{\mcitedefaultmidpunct}
{\mcitedefaultendpunct}{\mcitedefaultseppunct}\relax
\EndOfBibitem
\bibitem{Falk:1995th}
A.~F. Falk and T.~Mehen, \ifthenelse{\boolean{articletitles}}{{\it {Excited
  heavy mesons beyond leading order in the heavy quark expansion}},
  }{}\href{http://dx.doi.org/10.1103/PhysRevD.53.231}{Phys.\ Rev.\  {\bf D53}
  (1996) 231}, \href{http://arxiv.org/abs/hep-ph/9507311}{{\tt
  arXiv:hep-ph/9507311}}\relax
\mciteBstWouldAddEndPuncttrue
\mciteSetBstMidEndSepPunct{\mcitedefaultmidpunct}
{\mcitedefaultendpunct}{\mcitedefaultseppunct}\relax
\EndOfBibitem
\bibitem{Koponen:2007nr}
UKQCD collaboration, J.~Koponen, \ifthenelse{\boolean{articletitles}}{{\it
  {Energies of $B_s$ meson excited states: a lattice study}},
  }{}\href{http://dx.doi.org/10.1103/PhysRevD.78.074509}{Phys.\ Rev.\  {\bf
  D78} (2008) 074509}, \href{http://arxiv.org/abs/0708.2807}{{\tt
  arXiv:0708.2807}}\relax
\mciteBstWouldAddEndPuncttrue
\mciteSetBstMidEndSepPunct{\mcitedefaultmidpunct}
{\mcitedefaultendpunct}{\mcitedefaultseppunct}\relax
\EndOfBibitem
\bibitem{PDG}
Particle Data Group, J.~Beringer {\em et~al.},
  \ifthenelse{\boolean{articletitles}}{{\it {\href{http://pdg.lbl.gov/}{Review
  of particle physics}}},
  }{}\href{http://dx.doi.org/10.1103/PhysRevD.86.010001}{Phys.\ Rev.\  {\bf
  D86} (2012) 010001}\relax
\mciteBstWouldAddEndPuncttrue
\mciteSetBstMidEndSepPunct{\mcitedefaultmidpunct}
{\mcitedefaultendpunct}{\mcitedefaultseppunct}\relax
\EndOfBibitem
\bibitem{Aaltonen:2007ah}
CDF collaboration, T.~Aaltonen {\em et~al.},
  \ifthenelse{\boolean{articletitles}}{{\it {Observation of orbitally excited
  $B_s$ mesons}},
  }{}\href{http://dx.doi.org/10.1103/PhysRevLett.100.082001}{Phys.\ Rev.\
  Lett.\  {\bf 100} (2008) 082001}, \href{http://arxiv.org/abs/0710.4199}{{\tt
  arXiv:0710.4199}}\relax
\mciteBstWouldAddEndPuncttrue
\mciteSetBstMidEndSepPunct{\mcitedefaultmidpunct}
{\mcitedefaultendpunct}{\mcitedefaultseppunct}\relax
\EndOfBibitem
\bibitem{Abazov:2007af}
D0 collaboration, V.~Abazov {\em et~al.},
  \ifthenelse{\boolean{articletitles}}{{\it {Observation and properties of the
  orbitally excited $B_{s2}^*$ meson}},
  }{}\href{http://dx.doi.org/10.1103/PhysRevLett.100.082002}{Phys.\ Rev.\
  Lett.\  {\bf 100} (2008) 082002}, \href{http://arxiv.org/abs/0711.0319}{{\tt
  arXiv:0711.0319}}\relax
\mciteBstWouldAddEndPuncttrue
\mciteSetBstMidEndSepPunct{\mcitedefaultmidpunct}
{\mcitedefaultendpunct}{\mcitedefaultseppunct}\relax
\EndOfBibitem
\bibitem{Godfrey:1986wj}
S.~Godfrey and R.~Kokoski, \ifthenelse{\boolean{articletitles}}{{\it {The
  properties of P-wave mesons with one heavy quark}},
  }{}\href{http://dx.doi.org/10.1103/PhysRevD.43.1679}{Phys.\ Rev.\  {\bf D43}
  (1991) 1679}\relax
\mciteBstWouldAddEndPuncttrue
\mciteSetBstMidEndSepPunct{\mcitedefaultmidpunct}
{\mcitedefaultendpunct}{\mcitedefaultseppunct}\relax
\EndOfBibitem
\bibitem{Colangelo:2012xi}
P.~Colangelo, F.~De~Fazio, F.~Giannuzzi, and S.~Nicotri,
  \ifthenelse{\boolean{articletitles}}{{\it {New meson spectroscopy with open
  charm and beauty}},
  }{}\href{http://dx.doi.org/10.1103/PhysRevD.86.054024}{Phys.\ Rev.\  {\bf
  D86} (2012) 054024}, \href{http://arxiv.org/abs/1207.6940}{{\tt
  arXiv:1207.6940}}\relax
\mciteBstWouldAddEndPuncttrue
\mciteSetBstMidEndSepPunct{\mcitedefaultmidpunct}
{\mcitedefaultendpunct}{\mcitedefaultseppunct}\relax
\EndOfBibitem
\bibitem{Zhong:2008kd}
X.-H. Zhong and Q.~Zhao, \ifthenelse{\boolean{articletitles}}{{\it {Strong
  decays of heavy-light mesons in a chiral quark model}},
  }{}\href{http://dx.doi.org/10.1103/PhysRevD.78.014029}{Phys.\ Rev.\  {\bf
  D78} (2008) 014029}, \href{http://arxiv.org/abs/0803.2102}{{\tt
  arXiv:0803.2102}}\relax
\mciteBstWouldAddEndPuncttrue
\mciteSetBstMidEndSepPunct{\mcitedefaultmidpunct}
{\mcitedefaultendpunct}{\mcitedefaultseppunct}\relax
\EndOfBibitem
\bibitem{Wang:2012pf}
Z.-H. Wang, G.-L. Wang, H.-F. Fu, and Y.~Jiang,
  \ifthenelse{\boolean{articletitles}}{{\it {The strong decays of orbitally
  excited $B^{*}_{sJ}$ mesons by improved Bethe-Salpeter method}},
  }{}\href{http://dx.doi.org/10.1016/j.physletb.2011.11.051}{Phys.\ Lett.\
  {\bf B706} (2012) 389}, \href{http://arxiv.org/abs/1202.1224}{{\tt
  arXiv:1202.1224}}\relax
\mciteBstWouldAddEndPuncttrue
\mciteSetBstMidEndSepPunct{\mcitedefaultmidpunct}
{\mcitedefaultendpunct}{\mcitedefaultseppunct}\relax
\EndOfBibitem
\bibitem{Belle:2011aa}
Belle collaboration, A.~Bondar {\em et~al.},
  \ifthenelse{\boolean{articletitles}}{{\it {Observation of two charged
  bottomonium-like resonances in $\Upsilon(5S)$ decays}},
  }{}\href{http://dx.doi.org/10.1103/PhysRevLett.108.122001}{Phys.\ Rev.\
  Lett.\  {\bf 108} (2012) 122001}, \href{http://arxiv.org/abs/1110.2251}{{\tt
  arXiv:1110.2251}}\relax
\mciteBstWouldAddEndPuncttrue
\mciteSetBstMidEndSepPunct{\mcitedefaultmidpunct}
{\mcitedefaultendpunct}{\mcitedefaultseppunct}\relax
\EndOfBibitem
\bibitem{Adachi:2012cx}
Belle collaboration, I.~Adachi {\em et~al.},
  \ifthenelse{\boolean{articletitles}}{{\it {Study of three-body
  $\Upsilon(10860)$ decays}}, }{}\href{http://arxiv.org/abs/1209.6450}{{\tt
  arXiv:1209.6450}}\relax
\mciteBstWouldAddEndPuncttrue
\mciteSetBstMidEndSepPunct{\mcitedefaultmidpunct}
{\mcitedefaultendpunct}{\mcitedefaultseppunct}\relax
\EndOfBibitem
\bibitem{Bondar:2011ev}
A.~Bondar {\em et~al.}, \ifthenelse{\boolean{articletitles}}{{\it {Heavy quark
  spin structure in $Z_b$ resonances}},
  }{}\href{http://dx.doi.org/10.1103/PhysRevD.84.054010}{Phys.\ Rev.\  {\bf
  D84} (2011) 054010}, \href{http://arxiv.org/abs/1105.4473}{{\tt
  arXiv:1105.4473}}\relax
\mciteBstWouldAddEndPuncttrue
\mciteSetBstMidEndSepPunct{\mcitedefaultmidpunct}
{\mcitedefaultendpunct}{\mcitedefaultseppunct}\relax
\EndOfBibitem
\bibitem{Alves:2008zz}
LHCb collaboration, A.~A. Alves~Jr. {\em et~al.},
  \ifthenelse{\boolean{articletitles}}{{\it {The \lhcb detector at the LHC}},
  }{}\href{http://dx.doi.org/10.1088/1748-0221/3/08/S08005}{JINST {\bf 3}
  (2008) S08005}\relax
\mciteBstWouldAddEndPuncttrue
\mciteSetBstMidEndSepPunct{\mcitedefaultmidpunct}
{\mcitedefaultendpunct}{\mcitedefaultseppunct}\relax
\EndOfBibitem
\bibitem{Aaij:2012me}
R.~Aaij {\em et~al.}, \ifthenelse{\boolean{articletitles}}{{\it {The LHCb
  trigger and its performance}}, }{}\href{http://arxiv.org/abs/1211.3055}{{\tt
  arXiv:1211.3055}}\relax
\mciteBstWouldAddEndPuncttrue
\mciteSetBstMidEndSepPunct{\mcitedefaultmidpunct}
{\mcitedefaultendpunct}{\mcitedefaultseppunct}\relax
\EndOfBibitem
\bibitem{Sjostrand:2006za}
T.~Sj\"{o}strand, S.~Mrenna, and P.~Skands,
  \ifthenelse{\boolean{articletitles}}{{\it {PYTHIA 6.4 physics and manual}},
  }{}\href{http://dx.doi.org/10.1088/1126-6708/2006/05/026}{JHEP {\bf 05}
  (2006) 026}, \href{http://arxiv.org/abs/hep-ph/0603175}{{\tt
  arXiv:hep-ph/0603175}}\relax
\mciteBstWouldAddEndPuncttrue
\mciteSetBstMidEndSepPunct{\mcitedefaultmidpunct}
{\mcitedefaultendpunct}{\mcitedefaultseppunct}\relax
\EndOfBibitem
\bibitem{Belyaev:2011zza}
I.~Belyaev {\em et~al.}, \ifthenelse{\boolean{articletitles}}{{\it {Handling of
  the generation of primary events in \gauss, the \lhcb simulation framework}},
  }{}\href{http://dx.doi.org/10.1109/NSSMIC.2010.5873949}{Nuclear Science
  Symposium Conference Record (NSS/MIC) {\bf IEEE} (2010) 1155}\relax
\mciteBstWouldAddEndPuncttrue
\mciteSetBstMidEndSepPunct{\mcitedefaultmidpunct}
{\mcitedefaultendpunct}{\mcitedefaultseppunct}\relax
\EndOfBibitem
\bibitem{Lange:2001uf}
D.~J. Lange, \ifthenelse{\boolean{articletitles}}{{\it {The EvtGen particle
  decay simulation package}},
  }{}\href{http://dx.doi.org/10.1016/S0168-9002(01)00089-4}{Nucl.\ Instrum.\
  Meth.\  {\bf A462} (2001) 152}\relax
\mciteBstWouldAddEndPuncttrue
\mciteSetBstMidEndSepPunct{\mcitedefaultmidpunct}
{\mcitedefaultendpunct}{\mcitedefaultseppunct}\relax
\EndOfBibitem
\bibitem{Allison:2006ve}
GEANT4 collaboration, J.~Allison {\em et~al.},
  \ifthenelse{\boolean{articletitles}}{{\it {Geant4 developments and
  applications}}, }{}\href{http://dx.doi.org/10.1109/TNS.2006.869826}{IEEE
  Trans.\ Nucl.\ Sci.\  {\bf 53} (2006) 270}\relax
\mciteBstWouldAddEndPuncttrue
\mciteSetBstMidEndSepPunct{\mcitedefaultmidpunct}
{\mcitedefaultendpunct}{\mcitedefaultseppunct}\relax
\EndOfBibitem
\bibitem{Agostinelli:2002hh}
GEANT4 collaboration, S.~Agostinelli {\em et~al.},
  \ifthenelse{\boolean{articletitles}}{{\it {GEANT4: A simulation toolkit}},
  }{}\href{http://dx.doi.org/10.1016/S0168-9002(03)01368-8}{Nucl.\ Instrum.\
  Meth.\  {\bf A506} (2003) 250}\relax
\mciteBstWouldAddEndPuncttrue
\mciteSetBstMidEndSepPunct{\mcitedefaultmidpunct}
{\mcitedefaultendpunct}{\mcitedefaultseppunct}\relax
\EndOfBibitem
\bibitem{LHCb-PROC-2011-006}
M.~Clemencic {\em et~al.}, \ifthenelse{\boolean{articletitles}}{{\it {The \lhcb
  simulation application, \gauss: design, evolution and experience}},
  }{}\href{http://dx.doi.org/10.1088/1742-6596/331/3/032023}{{J.\ of Phys.\ :
  Conf.\ Ser.\ } {\bf 331} (2011) 032023}\relax
\mciteBstWouldAddEndPuncttrue
\mciteSetBstMidEndSepPunct{\mcitedefaultmidpunct}
{\mcitedefaultendpunct}{\mcitedefaultseppunct}\relax
\EndOfBibitem
\bibitem{Golonka:2005pn}
P.~Golonka and Z.~Was, \ifthenelse{\boolean{articletitles}}{{\it {PHOTOS Monte
  Carlo: A precision tool for QED corrections in $Z$ and $W$ decays}},
  }{}\href{http://dx.doi.org/10.1140/epjc/s2005-02396-4}{Eur.\ Phys.\ J.\  {\bf
  C45} (2006) 97}, \href{http://arxiv.org/abs/hep-ph/0506026}{{\tt
  arXiv:hep-ph/0506026}}\relax
\mciteBstWouldAddEndPuncttrue
\mciteSetBstMidEndSepPunct{\mcitedefaultmidpunct}
{\mcitedefaultendpunct}{\mcitedefaultseppunct}\relax
\EndOfBibitem
\bibitem{Breiman}
L.~Breiman, J.~H. Friedman, R.~A. Olshen, and C.~J. Stone, {\em Classification
  and regression trees}, Wadsworth international group, Belmont, California,
  USA, 1984\relax
\mciteBstWouldAddEndPuncttrue
\mciteSetBstMidEndSepPunct{\mcitedefaultmidpunct}
{\mcitedefaultendpunct}{\mcitedefaultseppunct}\relax
\EndOfBibitem
\bibitem{Roe}
B.~P. Roe {\em et~al.}, \ifthenelse{\boolean{articletitles}}{{\it {Boosted
  decision trees as an alternative to artificial neural networks for particle
  identification}},
  }{}\href{http://dx.doi.org/10.1016/j.nima.2004.12.018}{Nucl.\ Instrum.\
  Meth.\  {\bf A543} (2005) 577},
  \href{http://arxiv.org/abs/physics/0408124}{{\tt
  arXiv:physics/0408124}}\relax
\mciteBstWouldAddEndPuncttrue
\mciteSetBstMidEndSepPunct{\mcitedefaultmidpunct}
{\mcitedefaultendpunct}{\mcitedefaultseppunct}\relax
\EndOfBibitem
\bibitem{Pivk:2004ty}
M.~Pivk and F.~R. Le~Diberder, \ifthenelse{\boolean{articletitles}}{{\it
  {SPlot: A statistical tool to unfold data distributions}},
  }{}\href{http://dx.doi.org/10.1016/j.nima.2005.08.106}{Nucl.\ Instrum.\
  Meth.\  {\bf A555} (2005) 356},
  \href{http://arxiv.org/abs/physics/0402083}{{\tt
  arXiv:physics/0402083}}\relax
\mciteBstWouldAddEndPuncttrue
\mciteSetBstMidEndSepPunct{\mcitedefaultmidpunct}
{\mcitedefaultendpunct}{\mcitedefaultseppunct}\relax
\EndOfBibitem
\bibitem{Wu:1991xn}
Q.~Wu {\em et~al.}, \ifthenelse{\boolean{articletitles}}{{\it {Measurement of
  the $B^{*}$ cross-section at \sqs = 10.61 to 10.70~GeV}},
  }{}\href{http://dx.doi.org/10.1016/0370-2693(91)90574-A}{Phys.\ Lett.\  {\bf
  B273} (1991) 177}\relax
\mciteBstWouldAddEndPuncttrue
\mciteSetBstMidEndSepPunct{\mcitedefaultmidpunct}
{\mcitedefaultendpunct}{\mcitedefaultseppunct}\relax
\EndOfBibitem
\end{mcitethebibliography}

%\input{appendix}

%\bibliographystyle{unsrt}
%\bibliography{main}

%\clearpage
%\center{\bf{The LHCb Collaboration}}
%\onecolumngrid 
%\begin{widetext}
%\input{LHCb_authorlist}
%\end{widetext}
\end{document}